# Photonic Spike Processing: Ultrafast Laser Neurons and an Integrated Photonic Network

Bhavin J. Shastri*, Alexander N. Tait*, Mitchell A. Nahmias*, and Paul R. Prucnal†

Princeton University, Princeton, NJ, 08544 USA

* These authors contributed equally to this article
† To whom correspondence should be addressed. E-mail: prucnal@princeton.edu

**Abstract**—The marriage of two vibrant fields—photonics and neuromorphic processing—is fundamentally enabled by the strong analogies within the underlying physics between the dynamics of biological neurons and lasers, both of which can be understood within the framework of nonlinear dynamical systems theory. Whereas neuromorphic engineering exploits the biophysics of neuronal computation algorithms to provide a wide range of computing and signal processing applications, photonics offer an alternative approach to neuromorphic systems by exploiting the high speed, high bandwidth, and low crosstalk available to photonic interconnects which potentially grants the capacity for complex, ultrafast categorization and decision-making. Here we highlight some recent progress on this exciting field.

## Introduction

Recently, there has been a deeply committed exploration of unconventional computing paradigms (such as quantum [1] and neuro-inspired [2]) that promise to vastly outperform conventional technology in certain problem domains. While quantum computing exploits quantum-mechanical phenomena such as superposition and entanglement to study quantum coherent phenomena, neuromorphic engineering aims to build machines that better interact with natural environments by applying the circuit and system principles of neuronal computation, including robust analog signaling, physics-based dynamics, distributed complexity, and learning. Cognitive computing platforms inspired by the architecture of the brain promise potent advantages in efficiency, fault tolerance and adaptability over von Neumann architectures for tasks involving pattern analysis, decision making, optimization, learning, and real-time control of many-sensor, many-actuator systems. These neural-inspired systems are typified by a set of computational principles, including hybrid analog-digital signal representations, co-location of memory and processing, unsupervised statistical learning, and distributed representations of information.

The choice of information representation has strong implications on information processing. A sparse coding scheme called *spiking* has been recognized as a cortical encoding strategy with firm code-theoretic justifications [3] and the use of sparse coding principles promises extreme improvements to computational power efficiency [4]. Since spikes are discrete events that occur at analog times, this encoding scheme represents a hybrid between traditional analog and digital approaches, capable of both expressiveness and robustness to noise. A spike primitive handles inputs from multiple sources by temporally integrating their weighted sum and firing a single spike when this value crosses a threshold. These simple computational elements integrate a small set of basic operations (delay, weighting, spatial summation, temporal integration, and thresholding) into a single device, which is capable of performing a variety of computations depending on how its parameters are configured. This distributed, asynchronous model processes information using both space and time, and is amenable to algorithms for unsupervised adaptation [5].

## Electronics and Photonics

Spiking primitives have been built in CMOS analog circuits, digital "neurosynaptic cores", and non-CMOS devices. Various technologies have demonstrated large-scale spiking neural networks in electronics, including, notably: Neurogrid as part of Stanford University's Brains in Silicon program, IBM's TrueNorth as part of DARPA's SyNAPSE program, Spikey as part of University of Heidelberg's FACETS project, and University of Manchester's neuromorphic chip as part of the SpiNNaker project; the latter two are under the flagship of the European Commission's Human Brain Project. Unlike conventional processors, neuromorphic processors must achieve the massive fan-in essential for neural network architectures. Although some incorporate crossbar arrays—a dense mesh of wires overlaying the semiconductor substrate—all are ultimately forced to adopt some form of address-event routing (AER), which introduces the overhead of representing spike as digital codes instead of physical pulses. Generally, microelectronic architectures target biological time scales and the associated application spaces, and AER exploits the speed discrepancy between electronic circuitry and biological electrophysiology. AER systems are subject to the typical traffic restrictions of packet-routed networks, so neuromorphic processing for high-bandwidth applications must take a fundamentally different approach to interconnection. Just as photonics are being employed for interconnection in conventional CPU systems, optical networking principles can be applied to the neuromorphic domain.

The ability to map a processing paradigm to its underlying dynamics, rather than abstracting the physics away entirely, can significantly streamline efficiency and performance. The marriage of photonics with spike processing is fundamentally enabled by the strong analogy between the underlying physics biological neuron dynamics and lasers, both of which can be understood within the framework of dynamical systems theory. Integrated photonic platforms offer an alternative approach to microelectronics. The high switching speeds, high communication bandwidth, and low cross talk achievable in photonics are very well suited for an ultrafast spike-based information scheme with high interconnection densities. In addition, the high wall-plug efficiencies of photonic devices may allow such implementations to match or eclipse equivalent electronic systems in low energy usage. Because of these advantages, photonic spike processors could access a *picosecond, low-power* computationally rich domain that is inaccessible by other technologies. Our work aims to synergistically integrate the underlying physics of photonics with bio-inspired spike-based processing (Figure 1). This novel processing domain—*ultrafast cognitive computing*—represents a broad domain of applications where quick, temporally precise and robust systems are necessary, including: adaptive control, learning, perception, motion control, sensory processing, autonomous robotics, and cognitive processing of the radio frequency (RF) spectrum.

**Challenges**

The three key criteria for nonlinear elements to enable a practical scalable computing platform are: logic-level restoration, fanout/input-output isolation, and cascadability [6][7]. Past approaches to optical computing have met challenges realizing these requirements. We have hypothesized that a primary barrier to exploiting the high bandwidth of photonic devices for computing lies not in the performance, integration, or fabrication of devices, but instead in the model of computation being used (digital logic and specialized analog function acceleration). Analog photonic processing has found widespread application in high bandwidth filtering of microwave signals, but the accumulation of phase noise, in addition to amplitude noise, makes cascaded operations particularly difficult. Digital logic gates that suppress amplitude noise accumulation have also been realized in photonics, but proposed optical logic devices have not yet met all the conditions of cascadability, let alone scalability, which includes high fabrication yield. In addition, schemes that take advantage of the multiple available wavelengths require ubiquitous wavelength conversion, which can be costly, noisy, and inefficient.

The optical channel is highly expressive and correspondingly very sensitive to phase and frequency noise. Any proposal for a computational primitive must address the issue of practical cascadability, especially if multiple wavelength channels are intended to be used. Our unconventional computing primitive addresses the traditional problem of noise accumulation by interleaving physical representations of information. Representational interleaving, in which a signal is repeatedly transformed between coding schemes (digital-analog) or physical variables (electronic-optical), can grant many advantages to computation and noise properties. For example, a logarithm transform can reduce a multiplication operation to a simpler addition operation. The spiking model found in biology naturally interleaves robust, discrete representations for communication with precise, continuous representations for computation in order to reap the benefits of both types of coding. It is natural to deepen this distinction to include physical representational aspects, with the important result that optical noise does not accumulate. When a pulse is generated, it is transmitted and routed through a linear optical network with the help of its wavelength identifier. It is received only once into a fermionic degree of freedom, such as a deflection of carrier concentration in a photonic semiconductor or a current pulse in a photodiode. The soma tasks occur in a domain that is computationally richer (analog), yet more physically robust (incoherent) to the type of phase and frequency noise that can frustrate optical computing architectures.

Another major hurdle faced by substantial photonic systems is relatively poor device yield. While difficult for digital circuits to build in redundancy overhead, neuromorphic systems are naturally reconfigurable and adaptive. Biological nervous systems adapt to shunt signals around faulty neurons without the need for external control, a possible strategy for resistance to fabrication defects.

**Spiking Elements: Laser Neuron**

Recent years have seen the emergence of a new class of optical devices that exploit a dynamical isomorphism between semiconductor photocarriers and neuron biophysics. The difference in physical timescales allows these "photonic neurons" to exhibit spiking behavior on picosecond (instead of millisecond) timescales [8]. Spiking is closely related to a dynamical system property that underlies all-or-none responses called *excitability*, which is shared by certain kinds of laser devices. Excitable laser systems have been studied in the context of spike processing with the tools of bifurcation theory by [9] and experimentally by [10][11].

We recently designed excitable lasers specifically designed for compatibility with common photonic integrated circuit (PIC) platforms [12]. Specifically, at the 2013 IEEE Photonics Conference, we proposed [13] a hybrid silicon distributed feedback (DFB) laser and photodetector system that can emulate both a *leaky integrate-and-fire* (LIF) neuron and a synaptic variable, completing a computational paradigm that can be used to emulate a wide variety of functional cortical algorithms. Networks of such devices are easily scalable using a silicon III-V wafer-bonding platform [14]. The LIF neuron model is a mathematical model of the spiking dynamics which pervade animal nervous systems, well-established as the most widely used model of biological neurons in theoretical neuroscience for studying complex computation in nervous systems. A simple model of a single-mode laser with saturable absorber (SA) section has been proven to be analogous to the equations governing an LIF neuron in certain parameter regimes (Figure 2) [12][15].

As illustrated in Figure 2, our device consists of three primary components: two photodetectors and an excitable laser. The photodetectors receive optical pulses from a network and produce a push-pull current signal which modulates the laser carrier injection. The excitable laser acts as a threshold decision maker and clean pulse generator analogous to the neural axon hillock. We chose to implement this model using a hybrid silicon evanescent DFB laser [14], the device on the left in Figure 3(a).

*1) The Neural Front-End:* Inputs from other laser neurons are weighted in the optical domain before reaching the excitatory or inhibitory photodetector. Each photodetector produces a photocurrent summing the total optical power. Demultiplexing many input channels is not necessary because the incoherent sum of all wavelength-division multiplexed (WDM) channels within the InP detection band is intentionally computed by the photodetector. Photodetectors have been demonstrated on the same hybrid silicon evanescent platform used for laser gain sections with response times limited by parasitic capacitance [16].

Dynamics introduced by the photodetector are analogous to synaptic dynamics governing the concentration of neurotransmitters in between signaling biological neurons. Photocurrent flow in a real photodiode and neural synaptic dynamics can be modeled by a first-order low-pass filter. Photodetector time constants can be controlled with different biasing and device dimensions, and can be varied to obtain a large repertoire of processing behaviors.

Analog excitatory and inhibitory photocurrents are subtracted passively by a push-pull wire junction. The net photocurrent conducts over a short wire to modulate the laser gain section. These wires are roughly analogous to passive dendritic conduction, with the key difference that there are only two wires regardless of the number of input channels. The sensitivity-bandwidth of the neural front-end is not significantly reduced during electronic conversion and passive processing, although transmission lines can suffer from many effects that render them unsuitable for high-bandwidth interconnects. Distortions introduced by impedance mismatch, attenuation, dispersion, and radiative interference coupling are all negligible for wires much shorter than the signals of interest. We employ a co-integrated wire design to keep the electronic connection local. A 20 $\mu$m wire has a characteristic conduction delay of nearly 10 THz, and will not introduce significant transmission line distortion for sub-THz signals.

*2) The Excitable Laser:* A two-section DFB hybrid evanescent laser is used to implement excitable dynamics. The DFB cavity has a single longitudinal lasing mode and allows lithographic definition of lasing wavelength via silicon on insulator (SOI) grating pitch. A proton implantation region electrically isolates the semiconductor absorber and gain sections. Small modulation currents that exceed the *Q*-switch threshold will trigger large pulse discharges. We use a standard two-section laser rate equation to model the described device with realistic parameters based on those found in [12][14][16]. The results are shown in Figure 3(b).

**Current Experimental Data**

*1) Fiber-based excitable laser:* We recently demonstrated [11] a fiber-based excitable laser as a proof of concept of excitability with an embedded SA. While the performance of this benchtop prototype is much less than that of an integrated version (bandwidth: 100 kHz, energy per spike: 10 nJ), it experimentally confirms the possibility of using laser systems to emulate the spike processing capabilities required for cortical processing: temporal integration of multiple inputs, threshold detection, and all-or-nothing pulse generation. Figure 4 is a demonstration of the system's ability to exhibit excitability when a series of spikes are incident upon it. Excitatory pulses increase the gain carrier concentration, which performs temporal integration. Enough excitation results in an excursion from equilibrium causing the laser to fire a pulse due to the saturation of the absorber to transparency. This is followed by a relative *refractory period* during which an excitatory pulse is unable to cause the laser to fire. The phase-space excursion resulting from an excitable response is stereotyped and repeatable while subthreshold activation results in no output: key all-or- nothing properties for pulse regeneration, reshaping, and signal integrity.

*2) Signal feature recognition:* We had previously demonstrated [17] a device for signal feature recognition based on the escape response neuron model of a crayfish. Crayfish escape from danger by means of a rapid escape response behavior. The corresponding neural circuit is configured to respond to appropriately sudden stimuli. Since this corresponds to a life-or-death decision for the crayfish, it must be executed quickly and accurately. A potential application of the escape response circuit based on lightwave neuromorphic signal processing could be for pilot ejection from military aircraft. Our device, which mimics the crayfish circuit using photonic technology, is sufficiently fast to be applied to defense applications in which critical decisions need to be made quickly while minimizing the probability of false alarm.

A simple two-neuron anomaly detection circuit mimicking the tail-flip response of a crayfish was built to demonstrate the unprecedented signal feature recognition and stimulus response of these photonic devices [Figure 5(a) and (b)]. Our analog optical model exploits fast (subnanosecond) signal integration and ultrafast (picosecond) optical thresholding. The first neuron is configured to respond to a set of signals with specific features, while the second neuron further selects a subset of the signal from a set determined by a weighting and delay configuration and responds only when the input stimuli and the spike from the first neuron arrive within a very short time interval. Figure 5(c) shows the experimental measurements of the recognition circuit detecting specific input patterns having precise time intervals between the inputs as we configured.

The benchtop model provides a testbed for exploring the synergy between spike processing and optical physics. Microfabrication of the model will allow more complex processing with significant power reduction offered by integration.

**Spiking Architectures: Integrated Photonic Network**

The communication potentials of optical interconnects (bandwidth, energy use, electrical isolation) have received attention for neural networking in the past; however, attempts to realize holographic or matrix-vector multiplication systems have encountered practical barriers, largely because they cannot be integrated. Techniques in silicon PIC fabrication is driven by a tremendous demand for optical communication links within conventional supercomputing systems [18]. The first platforms for systems integration of active photonics are becoming commercial reality [14][16], and promise to bring the economies of integrated circuit manufacturing to optical systems. Our work investigates the potential of modern PIC platforms to enable large-scale all-optical systems for unconventional and/or analog computing, using a standard device set designed for digital communication (waveguides, filters, detectors, etc.).

At the 2014 IEEE Optical Interconnects Conference, we presented [19] an on-chip networking approach called "broadcast-and-weight" that could support massively parallel interconnection between photonic spiking neurons. Our scheme leverages recent advances in PIC technology to address interconnect challenges faced by distributed processing. It has similarities with the fiber networking technique broadcast-and-select, which channelizes usable bandwidth using WDM; however, the protocol flattens the traditional layered hierarchy of optical networks, accomplishing physical, logical, and processing tasks in a compact computational primitive. Although its processing circuits are unconventional, the required device set is compatible with mainstream PIC platforms. WDM effectively channelizes available bandwidth without spatial or holographic multiplexing and avoids coherent interference effects during fan-in. High-bandwidth optical channels are compatible with our proposed laser neuron devices, which could access a picosecond computational domain that impacts application areas where both complexity and speed are paramount (e.g. adaptive control, real-time embedded system analysis, and cognitive RF processing).

*1) Broadcast-and-weight:* Broadcast-and-select is a fiber WDM protocol that obtains collision-free, circuit-routed, and densely parallel interconnection. The active connection is selected, not by altering the intervening medium, but by tuning a filter at the receiver. We have proposed a similar protocol called "broadcast-and-weight," which allows multiple inputs to be selected simultaneously and with intermediate strengths between 0% and 100%. A group of nodes shares a common medium in which the output of every node is assigned a unique transmission wavelength for broadcast (Figure 6).

*2) Processing-network node:* Participants in the network called processing-network nodes (PNN) perform both roles of processing (weighting, addition, nonlinear dynamics) and networking (routing, $\lambda$-fan-in, WDM carrier generation) in a compact set of standard devices. A silicon photonic PNN implementation is depicted in Figure 2. This circuit technique could also generalize to future PIC platforms. $\lambda$-fan-in in a photodetector strips WDM signals of any trace of their origin, a side-effect that corrupts digital signals, typically necessitating demultiplexing and dedicated detection. In the neurocomputing context, however, this channel destruction is precisely correspondant with the summation function, so demultiplexing is not required. A $\lambda$-fan-in front-end was found to yield input commutativity resulting in combinatorial robustness to device failures. The systemic failure rate of a PNN network can *decrease* with network size, in some cases, potentially making the integrated system more reliable than its constituent devices (Figure 7). A "receiver-less" front-end, it is not subject to well-known optical-electronic-optical (O/E/O) conversion overhead, whose assumed cost, energy, and complexity are due to the digital electronic receiver (amplifier, sampler, quantizer) in most communication links, not to the physical conversion itself.

*3) Broadcast loop:* A physical medium is efficiently implemented by a ring waveguide (Figure 8). Each PNN drops a fraction of total power, allowing most to continue. Drop-and-continue is a physical solution to optical multicasting that can enhance virtual interconnect density for a given network traffic [20]. In this context, the broadcast loop (BL) is fully multiplexed and capable of supporting $N^2$ interconnects in just 1 link with $N$ WDM channels, where an electronic interconnect would require, at best, $N(N-1)/2$ links.

Waveguide rings with WDM channelization have previously been proposed as an implementation of broadcast-and-*select* for efficient multicast in multi-core networks on-chip [21]. However, demultiplexing and dedicated detection can negate area and energy savings and create a buffering bottleneck. In contrast, physical $\lambda$-fan-in through total power detection does not require demultiplexing or an electronic receiver. By addressing interconnect challenges of efficient parallelism and fan-in, broadcast-and-*weight* could extend the promise of silicon photonics to high-performance neuromorphic architectures.

**Concluding Thoughts**

Neuromorphic networks have experienced a surge of popularity over the last seven years, motivated in part by bringing computation closer to its underlying physics. A modern generation of hardware-based neuron-inspired systems attempt to exploit the efficiency and robustness of sparse codes called spikes. Spiking systems could address important problems in recognition, optimization, and analysis of high-dimensional environments. Spiking models are central to several recent projects in microelectronic neuromorphic hardware, which address these domains where von Neumann machines perform poorly. Scaling to larger numbers of neurons, however, requires prohibitively complex networks in which there is a fundamental density-bandwidth trade-off. Furthermore, there are a growing number of applications that require higher speeds and lower latencies that may be outside the abilities of the fastest electronic circuits, including processing of the RF spectrum or ultrafast control. Alternatively, photonic systems can potentially offer much higher bandwidths and lower energy usage than electronics, making the spike-based approach to information processing a perfect fit for the technology. Taking advantage of the ephemeral dynamics in photonic systems such as those in lasers could lead to processors that operate on picosecond time scales. In addition, there is a close analogy between the dynamics of lasers and those of biological systems, both of which can exhibit excitability.

The research highlights included herein have indicated the potential to design a hybrid silicon laser that emulates biological networks of neurons at ultrafast speeds. Silicon photonic platform development has revolved around point-to-point links for multi-core computing systems. We have examined an opportunity for this technology to extend to unconventional architectures that rely heavily on interconnect performance. Broadcast-and-weight is a new approach for joining neuron-inspired processing and optical interconnect physics. The LIF model with a synaptic variable, coupled with tunable routing in a passive SOI network on a scalable platform, could open computational domains that demand unprecedented temporal precision, power efficiency, and functional complexity, could potentially emulate a huge variety of cortical functions, including RF spectral awareness, adaptive control of many-antenna systems, and high dimensional feature extraction of input data.

**Acknowledgment**

This work was supported in part by Princeton University through the Pyne Fund and Essig Enright Fund for Engineering in Neuroscience and the National Science Foundation (NSF) Mid-InfraRed Technologies for Health and the Environment (MIRTHE) Center. The work of B.J.S was supported by the Banting Postdoctoral Fellowship administered by the Government of Canada through the Natural Sciences and Engineering Research Council of Canada (NSERC). The work of A.N.T and M.A.N was supported by the NSF Graduate Research Fellowship Program (GRFP).

**Figures**

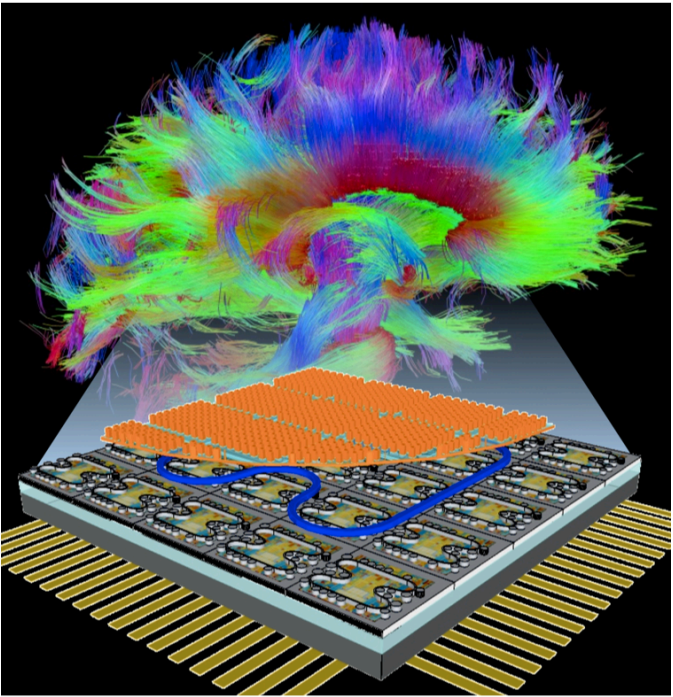

Figure 1. A conceptual rendering of a lightwave neuromorphic processor. Laser neuron arrays implement spiking dynamics with electo-optic physics, and a photonic network on-chip can support complex structures of virtual interconnection amongst these elements. Note: the colorful mind mapping (picture above the processor) was provided by the The Human Connectome Project.

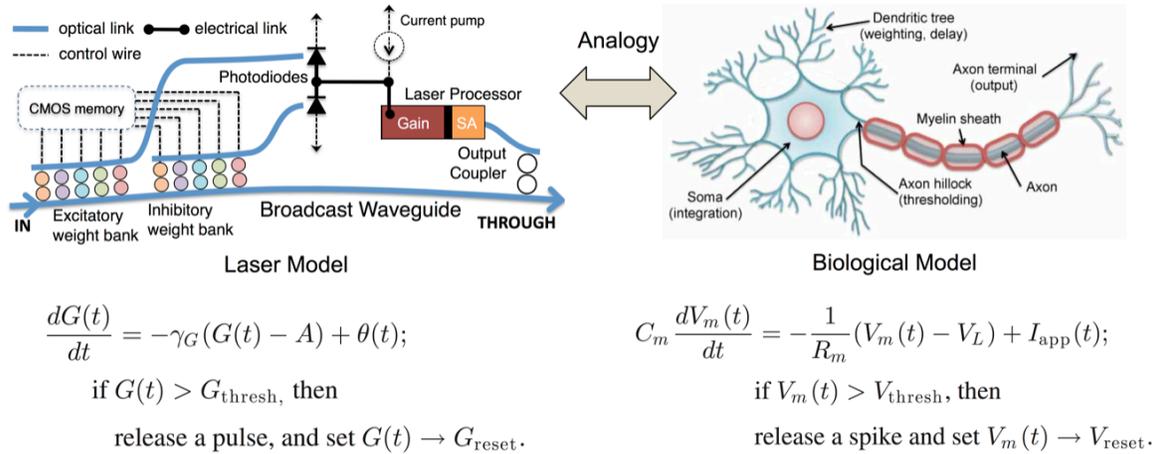

Figure 2. (Left) Implementation of the proposed device—processing-network node (PNN)—in standard silicon photonic devices. A spectral filter bank is implemented by a bank of tunable microring resonator filters, whose drop weight is configured by CMOS drivers. Balanced photodetectors sum all weighted inputs by converting to the electronic domain. Photocurrent subtraction allows inputs to have a positive (excitatory) or negative (inhibitory) modulation effect, push-pull capabilities considered essential to any neuron-inspired model. A short wire modulates current injection into a hybrid evanescent excitable laser neuron, which performs both threshold detection and pulse generation. The dynamics can be expressed with a simplified differential equation based on pulsed operation in the laser, where $G(t)$ models the gain, $\gamma_G$ is the relaxation rate of the gain, $A$ is the bias current of the gain, $\theta(t)$ represents input perturbations, $G_{\text{thresh}}$ is the gain threshold, and $G_{\text{reset}} \sim 0$ is the gain at transparency. The output of the laser is coupled back into the broadcast waveguide and broadcast to other PNNs. (Right) Analogy with a biological neuron. In the LIF neuron model, weighted and delayed input signals are spatially summed at the dendritic tree into an input current, $I_{\text{app}}(t)$, which travel to the soma and perturb the internal state variable, the voltage $V_m(t)$. The soma performs integration and then applies a threshold $V_{\text{thresh}}$, to make a spike or no-spike decision. After a spike is released, the voltage is reset $V_{\text{reset}}$. These dynamics can be encapsulated by the typical LIF model where $V_L$ is the membrane resting potential, and $R_m$ and $C_m$ model the resistance and capacitance associated with the membrane, respectively. The resulting spike is sent to other neurons in the network.

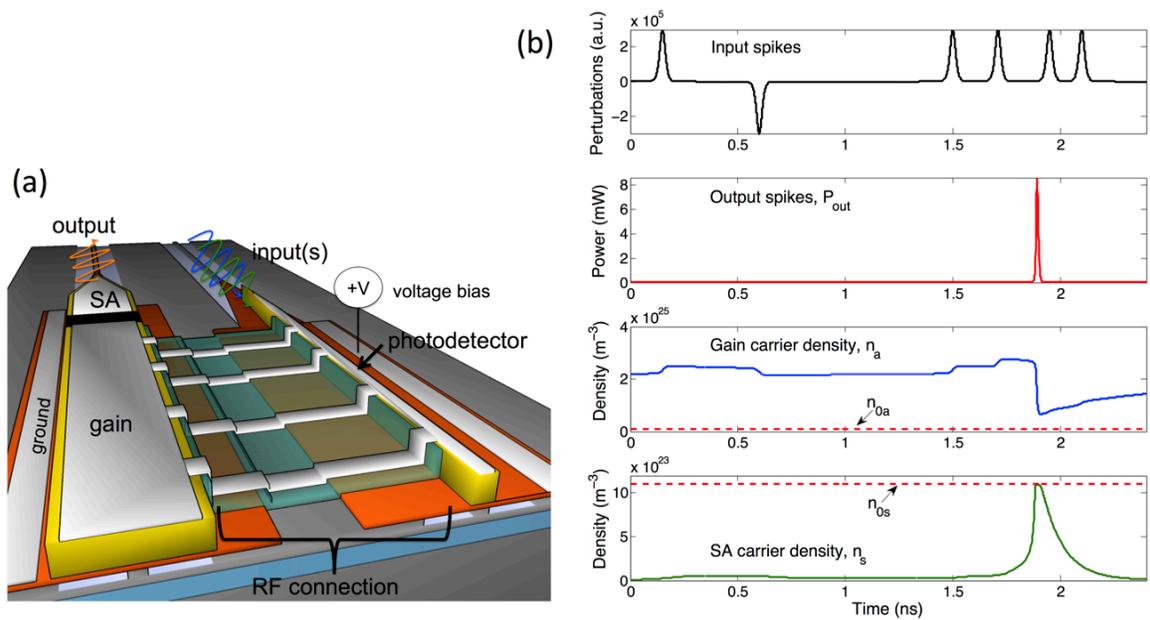

Figure 3. (a) Cross-section of a silicon hybrid evanescent laser neuron design. Optically active III-V components are bonded to an SOI substrate with rib waveguides. Optical spikes of potentially multiple wavelengths are incident onto the photodetector (right) from a passive SOI network. The resulting RF current pulse modulates the gain section of a two-section laser (left). (Inhibitory photodetector and pumping current source not shown for simplicity.) Legend: Si (gray), dielectric insulator (blue), III-V lower etch level (orange), III-V upper etch level (yellow), metal (white), proton implanted III-V (black). (b) A simulation of a DFB laser neuron with realistic parameters. Current perturbations from the photodetector (input spikes) modulate the gain (in blue). Enough excitatory activity causes the release of a pulse, followed by a short refractory period. This behavior mirrors that of a LIF neuron.

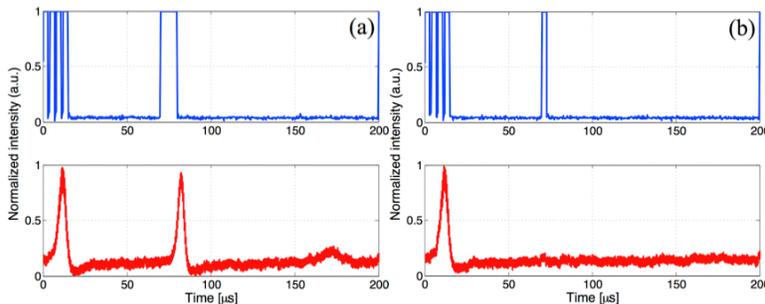

Figure 4. Experimental data of excitable fiber laser. (a) Traces showing a response to multiple integrated stimuli or one strong stimulus. (b) For too weak of an input, no change in output is detectable.

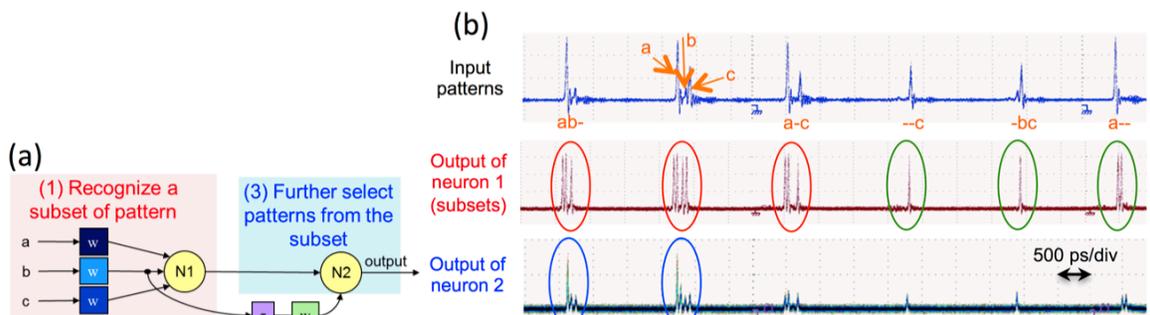

Figure 5. Experimental demonstration of crayfish response. (a) Laboratory setup. (b) Optical circuit diagram. The inputs $a$, $b$, and $c$ are weighted and delayed such that the first neuron is configured to spike for inputs with specific features. The neuron simultaneously normalizes and suppresses pulses above and below threshold, respectively. These spikes and part of input $b$ are launched into the second neuron. Through weighting and delaying of the inputs, spikes by the second integrator occur only for inputs with the desired features. The selection of the desired features can be reconfigured simply by adjusting the weights and delays of the inputs. (c)

Rudimentary ultrafast feature recognition. The top row shows various combinations of the three inputs with specific weights and delays. The input signals are integrated and thresholded at the output of neuron 1 (middle row). When the output spikes from the first neuron arrive at the second neuron slightly after input *b*, i.e., within the integration interval, the second neuron will spike. By adjusting the time delay of the inputs to neuron 2, the pattern recognizer identifies patterns *abc* and *ab−* (bottom row).

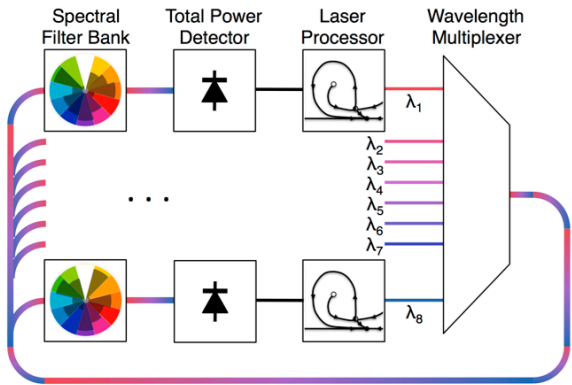

Figure 6. An optical broadcast-and-weight network showing parallels with the neural pathway model of weighted addition front-end and large fan-out. An array of source lasers are WDM in a single waveguide (multicolor). Independent weighting functions are realized by spectral filters (represented by gray color wheel masks) at the input of each unit. Instead of demultiplexing, the total optical power of each spectrally weighted signal is detected by a single photodetector, yielding the sum of the input channels. In this protocol, photodetectors act simultaneously as transducers and additive analog computational elements, solving both challenges of large, parallel fan-in and efficient many $\lambda$-conversion.

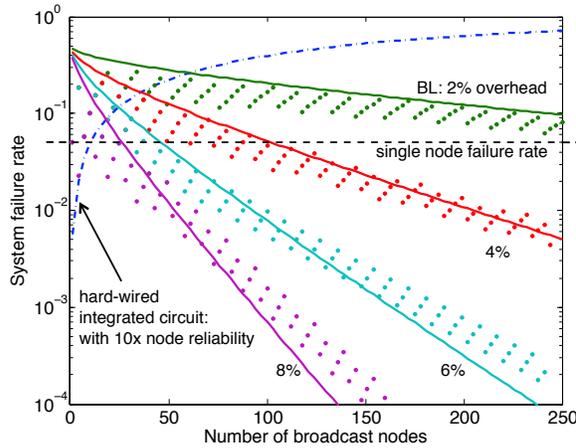

Figure 7. System failure rate in networks of unreliable nodes. In constrained circuit-routed networks (blue dash-dot line), system failure approaches certainty with more nodes. Broadcast-and-weight systems with hardware over- head (solid lines) invert this trend, experiencing reliability that scales with complexity. The systemic reliability of a BL can even be better, sometimes by orders of magnitude, than that of a single node (black dotted line). Colors represent different overhead percentages from 2% to 8%. Circular markers are the exact BL behavior; solid curves are approximate error function models.

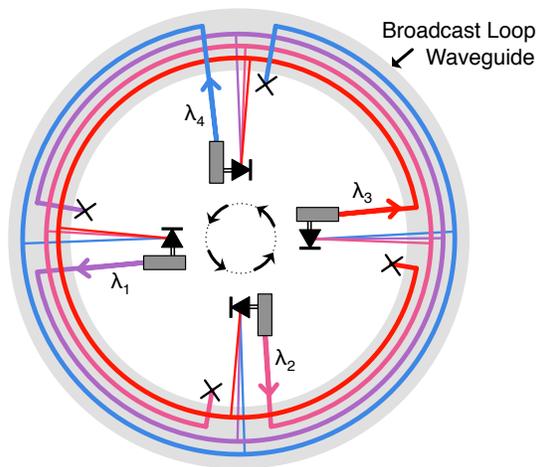

Figure 8. Conceptual diagram of a broadcast loop. The loop waveguide carries WDM channels from all participating PNNs. Each PNN detects a linear subset of the present channels. The PNN laser then outputs its signal, a function of those inputs, on its unique wavelength channel. Once a signal transverses the BL, it is terminated by its originating unit to avoid further interference. Filter banks and inhibitory pathways not shown.